\documentclass[twocolumn,aps,superscriptaddress]{revtex4}
\usepackage{graphicx}
\usepackage{amsmath}
\usepackage{amsfonts}
\usepackage{amssymb}
\usepackage{latexsym}
\usepackage{dcolumn}
\usepackage{jpc}

\begin{document}
\title{Dipole Formation at Interfaces of Alkanethiolate Self-assembled Monolayers and Ag(111)}
\author{Paul C. Rusu}
\affiliation{Computational Materials Science, Faculty of Science and Technology and MESA+ Institute
for Nanotechnology, University of Twente, P.O. Box 217, 7500 AE Enschede, The Netherlands}
\author{Gianluca Giovannetti}
\affiliation{Institute Lorentz for Theoretical Physics, Leiden University, P. O. Box 9506, 2300AE
Leiden, The Netherlands.} \affiliation{Computational Materials Science, Faculty of Science and
Technology and MESA+ Institute for Nanotechnology, University of Twente, P.O. Box 217, 7500 AE
Enschede, The Netherlands}
\author{Geert Brocks}
\altaffiliation[]{Corresponding author; phone: 31-53-489-3155; fax: 31-53-489- 2910; e-mail:
g.h.l.a.brocks@tnw.utwente.nl.} \affiliation{Computational Materials Science, Faculty of Science
and Technology and MESA+ Institute for Nanotechnology, University of Twente, P.O. Box 217, 7500 AE
Enschede, The Netherlands}
\date{\today}

\begin{abstract}
The formation of interface dipoles in self-assembled monolayers (SAMs) of --CH$_3$ and --CF$_3$
terminated short-chain alkanethiolates on Ag(111) is studied by means of density functional theory
calculations. The interface dipoles are characterized by monitoring the change in the surface work
function upon adsorption of the SAM. We compare results obtained for SAMs in structures with a
different packing density of molecules, i.e. \mbox{$(\sqrt{7}\times\sqrt{7}) R19.1^{\circ}$},
\mbox{$(\sqrt{3}\times\sqrt{3}) R30^{\circ}$}, and \mbox{p(2$\times$2)}. The work function of
alkanethiolate SAMs on silver depends weakly on the packing density; that of fluorinated
alkanethiolates shows a stronger dependance. The results are analyzed in terms of two nearly
independent contributions to the interface dipole. These originate respectively from the molecular
dipoles and from a charge transfer between the metal surface and the molecules. The charge transfer
is determined by the silver--sulfur bond and it is independent of the electronegativity of the
molecules.
\end{abstract}

\maketitle

\section{Introduction}

Self-assembled monolayers (SAMs) of organothiolates on metal surfaces are studied for a wide range
of technological applications running from catalysis, biosensors to microelectronic devices
\cite{Lee:lang03,Swalen:lang87,Chen:bp98}. In organic light-emitting diodes, the interfaces between
the metal contacts and the organic material are critical in the device performance, since they
control the injection of electrons and holes into the device \cite{Parker:app_phys94}.
Chemisorption of a SAM on a metal surface can alter its work function substantially. Depending on
the SAM, the work function can be manipulated advantageously to lower the energy barrier for
electron and hole injection \cite{Campbell:prb96,Campbell:apl97,de_Boer:adv_mat}.

Self-assembled monolayers have also become attractive for fundamental studies in metal--organic
interfaces and molecular electronics. They represent stable and ordered structures, which can be
prepared experimentally in air, in solution, or in vacuum \cite{Poirier:lang99,Schreiber:pss00}.
SAMs of alkanethiolates, C$_{n}$H$_{2n+1}$S, on Au(111) are among the most extensively studied
systems. Alkanethiolate SAMs on Au(111) adopt a \mbox{$(\sqrt{3}\times\sqrt{3}) R30^{\circ}$}
structure or superstructures thereof \cite{Vargas:jpcb01,Morikawa:ss02}. Alkanethiolates form SAMs
on a wide range of (noble) metal surfaces, which have a similar structure as on Au(111). Variations
on the packing density are possible, however, and on Ag(111) a somewhat denser packed
\mbox{$(\sqrt{7}\times\sqrt{7}) R19.1^{\circ}$} structure has been reported \cite{Schreiber:pss00}.

The change in work function of the surface upon adsorption of the SAM is directly proportional to
the dipole moment density generated at the SAM--metal interface. For SAMs on Au(111) it has been
shown that this dipole moment density is mainly determined by the permanent dipoles in the thiolate
molecular layer. The sulfur--gold bonds that are formed upon adsorption, are nearly apolar and give
a very small contribution to the interface dipole
\cite{Renzi:prl05,Bredas:prl06,Rousseau:jpcb06,Rusu:jpcb06}. However, a small sulfur--metal bond
dipole is typical of gold and the existence of a much larger bond dipole is indicated by
experiments of alkanethiolate SAMs on silver \cite{de_Boer:adv_mat}. In a previous computational
study on model structures we have shown that large bond dipoles can be formed in the adsorption of
SAMs on Ag and Pt surfaces \cite{Rusu:prb06}.

In this paper we study the interface dipole formation resulting from adsorption of SAMs on Ag(111)
by first-principles density functional theory (DFT) calculations. In particular we examine the
influence of the structure and the packing density of the molecules in the SAM. The
\mbox{$(\sqrt{7}\times\sqrt{7}) R19.1^{\circ}$} packing, which is observed experimentally for
alkanethiolate SAMs on Ag(111) \cite{Schreiber:pss00}, is our starting point. We consider several
low energy structures \cite{Rieley:lang99,Yu:jpcb06}. The results are compared to the
\mbox{$(\sqrt{3}\times\sqrt{3}) R30^{\circ}$} structure, where the surface area per adsorbed
molecule is 29\% larger, which is the most common structure of alkanethiolate SAMs on other noble
metal (111) surfaces. We also consider the less densely packed \mbox{p(2$\times$2)} structure,
which has a 71\% higher surface area per molecule. Fluorinated alkanethiolate SAMs on Au(111) can
be observed in this structure \cite{Schreiber:pss00}, and it might be possible that this structure
is also formed by such molecules adsorbed on Ag(111). We show that although the interface dipole
density is smaller for less densily packed structures, it is not simply proportional to the packing
density due to dielectric screening in the molecular layer.

The commonly used DFT functionals describe the formation of chemical bonds and the resulting charge
distribution very well, but they fail to capture the van der Waals interactions between the alkyl
chains. Van der Waals interactions are relatively unimportant in short chain alkanethiolates, which
is why we focus on the short chain thiolates CH$_{3}$S and C$_{2}$H$_{5}$S. To elucidate the
influence of the polarity of the molecules on the SAM$-$metal interface dipole, we also study the
fluorinated thiolates CF$_{3}$S and CF$_{3}$CH$_{2}$S. Since the directions of the dipole moment of
fluorinated and of nonfluorinated thiolates are roughly opposite, this leads to an obvious
difference in the interface dipole between SAMs of the two types of molecules. In addition,
fluorinated thiolates have a much higher electronegativity. One would expect that by varying the
relative electronegativity of the surface and the molecules, one can modify the electron transfer
between surface and molecules, which would give an additional contribution to the interface dipole.

In this paper we will show that increasing the electronegativity by fluorinating the alkyl tails
does however not lead to a change in charge transfer. We will arrive at this conclusion by
analyzing the interface dipole and separating it into a contribution from the molecular dipoles
and from the charge reordering at the metal$-$SAM interface. By comparing these results to those
obtained for SAM$-$Au(111) and SAM$-$Pt(111) interfaces, it can be concluded that the charge
transfer depends on the metal surface and the nature of the sulfur$-$metal bond, but not on the
molecular tails.

The paper is organized as follows. In the next section we describe the techniques we use for
calculating and analyzing the interface dipoles and give details on the parameters used in the
calculations. Subsequently the results on the SAM$-$Ag(111) interfaces are discussed. First we
discuss the possible structures and then we analyze the interface dipoles. The last section
contains a short summary and the conclusions.

\section{Theoretical section}

\subsection{Total energy calculations}
The Ag(111) metal surface is represented by a slab of layers of metal atoms stacked according to an
\textit{fcc} ABC sequence. A typical slab consists of four layers. The SAM is adsorbed on one side
of the slab. The surface unit cell depends upon the monolayer structure and coverage. The cells
used in our calculations are \mbox{$(\sqrt{7}\times\sqrt{7}) R19.1^{\circ}$},
\mbox{$(\sqrt{3}\times\sqrt{3}) R30^{\circ}$}, and \mbox{p(2$\times$2)}, which contain 7, 3 and 4
metal atoms per layer, respectively. Periodic boundary conditions are applied in all three
directions. This means that not only are the cells repeated along the surface (the $xy$-plane), but
also the slabs are repeated in the $z$-direction. The atoms in neighboring cells are separated
along the $z$-direction by a vacuum region of $\sim 12$ \AA . To cancel the artificial interaction
between the dipoles of the repeated slabs, the Neugebauer-Scheffler dipole correction is applied
\cite{Neugebauer:prb92}.

The electronic structure is treated within density functional theory (DFT)
\cite{Hohenberg_phys_rev} using the PW91 functional \cite{Perdew:prb92} to describe the electronic
exchange and correlation. The calculations are performed with the program VASP (Vienna \textit{ab
initio} simulation package) \cite{Kresse:prb93, Kresse:prb96} using the projector augmented wave
(PAW) method \cite{Kresse:prb99, Bloechl:prb94}. For noble metal atoms the outer shell s and d
electrons are treated as valence electrons, and for first and second row elements the outer shell s
and p electrons. The valence pseudo wave functions are expanded in a basis set consisting of plane
waves. All plane waves up to a kinetic energy cutoff of 500 eV are included.

The geometries are optimized by allowing the atoms in the top two metal layers and the atoms in the
SAMs to relax. The $(1\times 1)$ surface unit cell parameter is fixed at its optimized bulk value
of 2.93 \AA. The calculations use a $\mathbf{k}$-point sampling mesh of $7\times7$ for the $(\sqrt
7 \times \sqrt 7)$ structure, $11\times11$ for the $(\sqrt 3 \times \sqrt 3)$ and $9\times 9$ for
the $(2 \times 2)$ structures, according to the Monkhorst-Pack scheme. For geometry optimization a
Methfessel-Paxton smearing is used with a broadening parameter of 0.2 eV \cite{Met_Pax:prb89}. The
energies of the optimized geometries are recalculated using the tetrahedron scheme
\cite{Bloechl:prb94a}. Tests regarding the slab thickness, vacuum thickness, $\mathbf{k}$-point
sampling grid, and plane wave kinetic energy cutoff are performed, from which we estimate that
total energy differences are converged to within $\sim 0.05$ eV.

The adsorption energy of the SAM is calculated by comparing the total energy of the slab (with the
adsorbed SAM), with that of the clean slab (with the top surface in its relaxed Ag(111) structure),
and the free alkanethiolate (radical) molecules. If SAM adsorption results in a reconstruction of
the surface that involves metal adatoms \cite{Yu:jpcb06,He:prb06}, we assume that these adatoms are
supplied by the bulk metal. The adsorption energy per molecule $E_\mathrm{ads}$ associated with a
surface structure that contains $M$ molecules, $N_\mathrm{s}$ metal atoms per layer and
$N_\mathrm{ad}$ adatoms is then given by
\begin{equation}
E_\mathrm{ads} = \frac{1}{M}\left[E_\mathrm{slab} - N_\mathrm{s}E_\mathrm{clean} -
N_\mathrm{ad}E_\mathrm{bulk}\right] - E_\mathrm{mol}, \label{eq:abs}
\end{equation}
where $E_\mathrm{slab}$ is the total energy of the slab, $E_\mathrm{clean}$ is the total energy of
the clean slab per surface atom (top surface in optimized Ag(111) structure), $E_\mathrm{bulk}$ is
the total energy of bulk Ag per atom, and $E_\mathrm{mol}$ is the total energy of an alkanethiolate
molecule. Note that $E_\mathrm{ads}$ is negative if the adsorption is exothermic.

To analyze the results we have also calculated several properties of isolated thiolate molecules:
dipole moments, ionization potentials, electron affinities and electronegativities. For these
calculations we use the GAMESS program \cite{Schmidt:jcc93}, and treat the electronic structure
within DFT using the B3LYP functional \cite{Becke:jcp93,Stephens:jpch94}. We use the aug-cc-pVTZ
basis set. Calculations with the smaller 6-311G** basis set give dipole moments that are up to
$\sim 0.15$ D smaller, and ionization potentials, electron affinities that differ by $\sim 0.1$ eV.

\subsection{Work functions and interface dipoles}
Interface dipoles can be extracted from the change in the surface work function upon adsorption of
a SAM on a metal surface, as will be described below. Surface work functions are evaluated from the
expression:
\begin{equation}
W = V(\infty)-E_{F}, \label{eq:1}
\end{equation}
where $V(\infty)$ is the electrostatic potential in vacuum and $E_{F}$ is the Fermi energy of the
bulk metal. $V(\infty)$ is extracted by calculating the average electrostatic potential in the
$xy$-planes of the slab:
\begin{equation}
\overline{V}(z) = \frac{1}{C}\iint_{cell}V(x,y,z)dxdy , \label{eq:2}
\end{equation}
where $C$ is the area of the surface unit cell and $V(x,y,z)$ is the total electrostatic potential.
The latter is generated on an equidistant real space grid and the integral is obtained by
straighforward numerical integration. In practice $\overline{V}(z)$ reaches an asymtotic value
$V(\infty)$ within a distance of $\sim 5$\AA{} from the surface into the vacuum.

\begin{figure}[!tbp]
\includegraphics[scale=0.4,clip=true]{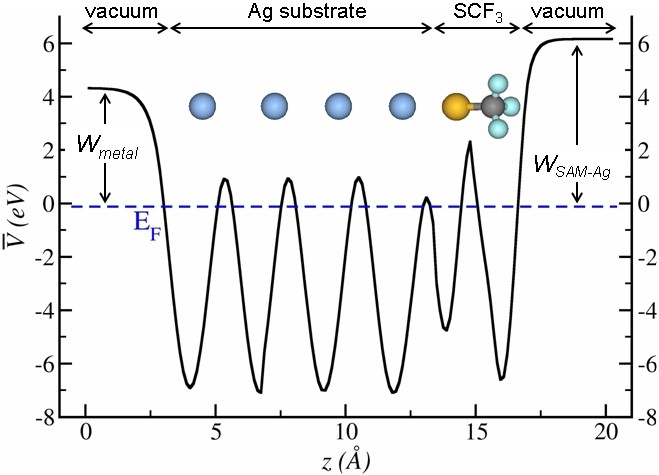}
\caption{Plane averaged electrostatic potential $\overline{V}(z)$ of a slab consisting of four
layers of silver atoms and one layer of CF$_{3}$S molecules on top. The $z$ axis is normal to the
(111) surface.} \label{work functiom}
\end{figure}

An example of $\overline{V}(z)$ is shown in Fig.~\ref{work functiom} for CF$_{3}$S on the Ag(111)
surface. $W_\mathrm{metal}$ is the work function of the clean Ag surface and $W_\mathrm{SAM-Ag}$ is
the work function of surface covered by the SAM. Slab calculations produce a reasonable value for
the bulk Fermi energy $E_{F}$, but a more accurate value is obtained from a separate bulk
calculation. We follow the procedure described by Fall \textit{et al.} \cite{Fall:jp99}. From the
convergence tests discussed in the previous section we estimate that calculated work functions are
converged to within $\sim 0.05$ eV. Typically, DFT/PW91 calculations give work functions that are
within 0.1-0.2 eV of the experimental values, although occasionally larger deviations of $\sim 0.3$
eV can be found \cite{Michaelides:prl03,Favot:chem_phys}.

Upon adsorption of a SAM the work function of a metal surface usually changes considerably. The
work function change $\Delta W$ can be interpreted in terms of a change in the surface dipole
$\Delta \mu$:
\begin{equation}
\Delta W = W_\mathrm{SAM-metal}-W_\mathrm{metal}=\dfrac{e \Delta \mu}{\varepsilon_{0}A},
\label{eq:3}
\end{equation}
where $A$ is the surface area per adsorbed molecule \cite{Jackson:book}. Note that $\Delta \mu$
corresponds to the component of the dipole moment directed along the surface normal, since only
this component affects the work function. $\Delta \mu$ is the result of the interface formation
between the SAM and the metal surface, and we call it the \textit{interface dipole} in the
following.

We split the interface dipole into a contribution $\mu_\mathrm{SAM}$ from the molecular dipoles in
the SAM, and a contribution $\mu_\mathrm{chem}$ from the charge transfer between the metal surface
and the molecule, which occurs upon chemisorption of the SAM. The latter contribution is then
defined by:
\begin{equation}
\mu_\mathrm{chem} = \Delta \mu - \mu_\mathrm{SAM}. \label{eq:4}
\end{equation}
$\mu_\mathrm{SAM}$ is obtained from a separate calculation on a free-standing SAM without the
presence of a metal slab, but with the molecules frozen in their adsorbed geometry. In the
following we will show that $\mu_\mathrm{chem}$ is nearly independent of the (fluorinated) alkyl
tail of the thiolate molecule. This means that $\mu_\mathrm{chem}$ is mainly determined by the
sulfur-metal bond and the charge transfer associated with this bond.

Note that the calculation of $\mu_\mathrm{SAM}$ is done for a full monolayer. In practice
$\mu_\mathrm{SAM}$ is obtained from the expression:
\begin{equation}
\mu_\mathrm{SAM}=\frac{\varepsilon_{0}A\Delta V}{e},  \label{eq:6}
\end{equation}
where $\Delta V = V(\infty) - V(-\infty)$ is the potential drop over the SAM, and $V(\infty),
V(-\infty)$ are the asymptotic electrostatic potentials on both sides of the SAM. These are easily
obtained, since the potential reaches its asymptotic values within a distance of few \AA{} of the
SAM.

This calculation incorporates the effect of the depolarizing electric field within the SAM that is
generated by the close-packed molecular dipoles. Often this effect is modeled phenomenologically by
introducing an effective dielectric constant $\varepsilon$ for the SAM:
\begin{equation}
\mu_{SAM} = \frac{\mu_z}{\varepsilon}, \label{eq:5}
\end{equation}
where $\mu_z$ is the $z$-component of the permanent dipole of the isolated molecule. By obtaining
$\mu_z$ from a separate calculation we will extract the effective dielectric constant as a function
of the packing density of the molecules in the SAM.

\section{Influence of packing density}

We will first discuss the possible structures of thiolate SAMs on the Ag(111) surface and then
study the influence of the packing density on the work function.

\subsection{Structures}
\begin{figure}[!tbp]

\includegraphics[scale=0.46,clip=true]{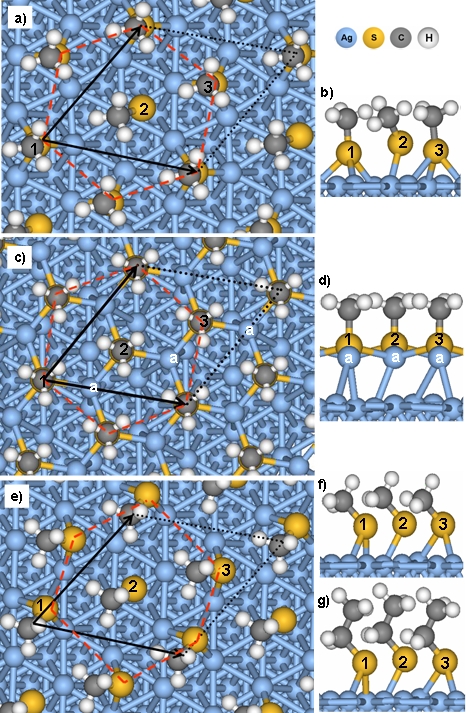}

\caption{Possible \mbox{$(\sqrt{7}\times\sqrt{7}) R19.1^{\circ}$} structures of CH$_3$S SAMs on
Ag(111); (a), (b) top and side view of the 1,3 hollow structure; (c), (d) of the reconstructed
structure; (e), (f) of the bridge structure; (g) of a C$_2$H$_5$S SAM in the bridge structure. ``1,
2, 3'' label the molecules, ``a'' labels the Ag adatoms. } \label{7times7structure}

\end{figure}

From early scanning tunneling microscopy (STM) experiments it was concluded that SAMs of
alkanethiolates on the Ag(111) surface form a commensurate \mbox{$(\sqrt{7}\times\sqrt{7})
R10.9^{\circ}$} structure \cite{Dhirani:lang95,Heinz:langm95}. Normal incident X-ray standing wave
(NIXSW) experiments have confirmed the \mbox{$\sqrt{7}\times\sqrt{7}$} structure, but have
corrected the registry of the SAM on the underlying substrate to \mbox{$(\sqrt{7}\times\sqrt{7})
R19.1^{\circ}$} \cite{Rieley:lang99}. The proposed model of this structure has the molecules in the
SAM arranged in a hexagonal lattice with a nearest neighbor distance between the sulfur atoms of
4.41 \AA, see Fig.~\ref{7times7structure}(a). Long chain alkanethiolates adopt an expanded
incommensurate \mbox{$(\sqrt{7}\times\sqrt{7}) R19.1^{\circ}$} structure with a nearest neighbor
distance of 4.6-4.8 \AA\ \cite{Fenter:lang91,Fonticelli:PhysChem2004}, whereas short chain
alkanethiolates keep the commensurate structure \cite{Rieley:lang99}. Recently a new model has been
proposed for the \mbox{$(\sqrt{7}\times\sqrt{7}) R19.1^{\circ}$} structure of CH$_{3}$S SAMs on
Ag(111)on the basis of NIXSW and medium energy ion scattering (MEIS) experiments
\cite{Yu:lang05,Yu:jpcb06,Parkinson:ss07}. It involves a surface reconstruction consisting of a 3/7
monolayer of Ag adatoms, which are bonded to the the methylthiolate molecules,
Fig.~\ref{7times7structure}(c).

The \mbox{$(\sqrt{7}\times\sqrt{7}) R19.1^{\circ}$} structure proposed first consists of three
molecules per surface unit cell with two of the molecules adsorbed on hollow sites and one on a top
site \cite{Rieley:lang99}. Starting from this structure we have relaxed the geometry of CH$_{3}$S
on Ag(111) and the result is shown in Fig.~\ref{7times7structure}(a) and (b). The molecules labeled
1 and 3 change their position only slightly and remain adsorbed in a hollow site. The molecule
labeled 2 moves away from the top site towards a bridge site. The angle between the surface normal
and the C--S bond is 42$^{\circ}$, whereas that angle for molecules 1 and 3 is only 9-10$^{\circ}$.
The latter molecules are almost standing upright, as can be observed in
Fig.~\ref{7times7structure}(b). We call this structure the ``1,3 hollow'' structure in the
following. The most important bond distances and angles of this structure are given in
Table~\ref{7times7data}. The adsorption energy (averaged, per molecule) according to
Eq.~(\ref{eq:abs}) is $-1.97$ eV.

In our previous calculations for alkanethiolate SAMs on Au(111) the molecules show a strong
preference for adsorption on bridge sites, instead of on hollow or top sites \cite{Rusu:jpcb06}.
Starting with CH$_{3}$S molecules 1 and 3 on bridge positions we obtain the optimized geometry that
is shown in Figs.~\ref{7times7structure}(e) and (f). In this structure the CH$_{3}$S molecule 2
also moves closer to the bridge position, as compared to the 1,3 hollow structure, see
Table~\ref{7times7data}. The new structure, which we call the ``bridge'' structure, is 0.10
eV/molecule lower in energy than the 1,3 hollow structure. The bridge structure is energetically
favored over the 1,3 hollow structure. The calculated adsorption energy in the bridge structure is
$-2.07$ eV/molecule. The geometries of the three molecules in the bridge structure are more similar
than in the 1,3 hollow structure. For instance, the angle between the surface normal and the C--S
bond is in the range 43-51$^{\circ}$ for all three molecules, see Table~\ref{7times7data}.  The
work functions of the 1,3 hollow and the bridge structures are substantially different as will be
discussed below.

We have also optimized the geometry in the  \mbox{$(\sqrt{7}\times\sqrt{7}) R19.1^{\circ}$}
reconstructed structure \cite{Yu:jpcb06,Parkinson:ss07}. The reconstruction involves a commensurate
layer of Ag adatoms at a 3/7 monolayer coverage. The sulfur atoms of the adsorbed thiolate
molecules are threefold coordinated by adatoms. The optimized geometry is shown in
Figs.~\ref{7times7structure}(c) and (d) and bond distances and angles of this structure are given
in Table~\ref{7times7data}. The resulting structure has the CH$_{3}$S molecules standing upright
with the C--S bond pointing along the surface normal in agreement with previous calculations
\cite{He:prb06,Torres:prl06}. The distance of the S atoms of the different molecules to the surface
is the same within $\sim 0.1$ \AA, which represents a very small ``rumpling'' in agreement with the
latest experimental results \cite{Parkinson:ss07}. The work function of the reconstructed structure
is substantially different from that of the unreconstructed structures as will be discussed below.
The calculated adsorption energy in the reconstructed structure is $-2.07$ eV/molecule. This number
is very close to the adsorption energy in the (unreconstructed) bridge structure. Within the
intrinsic error bar of DFT calculations the two structures are degenerate in energy. It has
recently been suggested that the two structures, i.e. reconstructed and unreconstructed, might
coexist on the surface \cite{Torres:prl06}.

The optimized unreconstructed \mbox{$(\sqrt{7}\times\sqrt{7}) R19.1^{\circ}$} structure of a
CH$_{3}$CH$_{2}$S SAM also has the molecules adsorbed on or near bridge sites, as shown in
Fig.~\ref{7times7structure}(g). The three molecules in the unit cell have a similar geometry. For
instance, the C--S--normal angle is 42-45$^{\circ}$, see Table~\ref{7times7ethyl}, which is similar
to the experimental value reported for CH$_{3}$(CH$_{2}$)$_7$S on Ag(111) \cite{Rieley:lang99}. The
chain angles are in the range 12-17$^{\circ}$, which is similar to experimental results for
long-chain alkanethiolates \cite{Laibinis:AmChemSoc1991}.

\begin{figure}[!tbp]

\includegraphics[scale=0.45,clip=true]{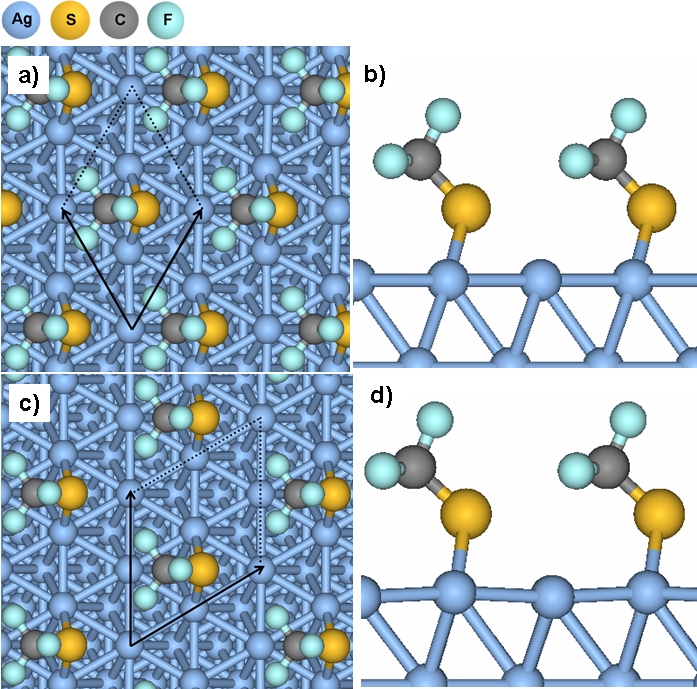}

\caption{CF$_{3}$S SAM on Ag(111) with the molecules adsorbed on bridge sites; (a),(b) top and side
view of the \mbox{$(\sqrt{3}\times\sqrt{3})R30^{\circ}$} structure; (c),(d) of the p(2$\times$2)
structure.} \label{SCF3}

\end{figure}

We have also studied \mbox{$(\sqrt{3}\times\sqrt{3})R30^{\circ}$} and p(2$\times$2) structures,
where the surface area per adsorbed molecule is 29\% and 71\% larger, respectively, see
Fig.~\ref{SCF3}. Experimentally it is not likely that alkanethiolates on Ag(111) form these
structures, but they enable us to model the influence of the packing density of the molecules in
the SAM on the work function and the interface dipoles. In both these structure the bridge site is
the favored adsorption site and the local geometry of the molecules is similar to that in the
\mbox{$(\sqrt{7}\times\sqrt{7}) R19.1^{\circ}$} bridge structure.

\begin{table}[btp]
\begin{tabular}{ l  c  c  c   }
\\
\hline
 & & 1,3 hollow &\\
\cline{2-4}
 &   molecule 1  &  2 & 3 \\
\hline
C--H(\AA) & 1.10 & 1.10 & 1.10\\
S--C(\AA) & 1.83 & 1.84 & 1.84\\
Ag--S(\AA)  & 2.52/2.50/2.50 & 2.47/3.01 & 2.63/2.54/2.54\\
C--S--normal$(^\circ)$ & 9.0 & 42.0 & 9.7\\
Ag--S--Ag$(^\circ)$ & 78.7/77.5/77.4 & - & 76.2/71.3/72.5 \\
\hline
& & bridge &\\
\cline{2-4}
C--H(\AA) & 1.10 & 1.10 & 1.10\\
S--C(\AA) & 1.83 & 1.84 & 1.83\\
Ag--S(\AA)  & 2.56/2.49 & 2.46/2.82 & 2.58/2.48\\
C--S--normal$(^\circ)$ & 51.4 & 43.0 & 46.6\\
Ag--S--Ag$(^\circ)$ & 72.3 & - & 72.5 \\
\hline
& & reconstructed &\\
 \cline{2-4}
C--H(\AA) & 1.10 & 1.10 & 1.10\\
S--C(\AA) & 1.84 & 1.84 & 1.84\\
Ag--S(\AA)  & 2.65 & 2.66 & 2.64\\
C--S--normal$(^\circ)$ & 0 & 0 & 0\\
Ag--S--Ag$(^\circ)$ & 114.9 & 115.3 & 115.9 \\
$d_z$& 2.47 & 2.56 & 2.47 \\
\end{tabular}
\caption{Bond lengths and bond angles of the \mbox{$(\sqrt{7}\times\sqrt{7}) R19.1^{\circ}$} 1,3
hollow, bridge and reconstructed structures of CH$_{3}$S SAMs on Ag(111). The columns indicate the
three molecules in the supercell, see Fig.~\ref{7times7structure}. $d_z$ is the distance along the
surface normal between a Ag adatom and the top Ag layer. } \label{7times7data}
\end{table}

\begin{table}[btp]
\begin{tabular}{ l  c  c  c   }
\\
\hline
&   molecule 1  &  2 & 3 \\
\hline
C--H(\AA) & 1.10 & 1.10 & 1.10\\
C--C(\AA) & 1.52 & 1.53 & 1.52\\
S--C(\AA) & 1.85 & 1.85 & 1.84\\
Ag--S(\AA)  & 2.57/2.50 & 2.47/2.90 & 2.58/2.49\\
C--C--normal$(^\circ)$ & 25.5 & 26.3 & 26.8\\
C--S--normal$(^\circ)$ & 44.2 & 41.9 & 45.0\\
chain$(^\circ)$ & 14.7 & 12.0 & 17.1\\
Ag--S--Ag$(^\circ)$ & 72.1 & - & 72.5 \\
\end{tabular}
\caption{Bond lengths and bond angles of the \mbox{$(\sqrt{7}\times\sqrt{7}) R19.1^{\circ}$} bridge
structure of CH$_{3}$CH$_{2}$S SAMs on Ag(111). The chain angle represents the angle made by the
line connecting the top C and S atoms with the surface normal.} \label{7times7ethyl}
\end{table}

The structure of (partially) fluorinated alkanethiolates is much less established than that of
their nonfluorinated counterparts. On Au(111) thiolates with long fluorinated alkyl tails have a
less dense packing because of their relatively bulky tails \cite{Schreiber:pss00,Liu:JCP94}. For
such SAMs a p(2$\times$2) structure has been proposed, where the spacing between the adsorbate
molecules is \mbox{5.87 \AA{}} \cite{Carla:lang93}. SAMs of long-chain alkanethiolates with
fluorinated end groups on Au(111) have a \mbox{$(\sqrt{3}\times\sqrt{3})R30^{\circ}$} structure
\cite{Pflaum:surf02}. The spacing between the adsorbate molecules is then \mbox{5.08 \AA{}}.

We did not find reports on the structural details of fluorinated alkanethiolate SAMs on silver in
the literature. We optimized the structure of CF$_3$S and CF$_3$CH$_2$S SAMs on Ag(111) in three
different packing densities, i.e., \mbox{$(\sqrt{7}\times\sqrt{7})$},
\mbox{$(\sqrt{3}\times\sqrt{3})$} and p(2$\times$2). For the \mbox{$(\sqrt{7}\times\sqrt{7})
R19.1^{\circ}$} structure of a CF$_3$S SAM, which represents the most dense packing, we used the
1,3 hollow and the bridge structures of CH$_3$S as starting points and optimized the geometry. The
bridge structure of CF$_3$S is more stable than the 1,3 hollow structure, albeit by less than 0.02
eV/molecule. However, even in the bridge structure only molecules 1 and 3 are actually adsorbed on
bridge sites, whereas molecule 2 is adsorbed on a hollow site. This results in a distorted
hexagonal packing of the CF$_3$S molecules.

For the \mbox{$(\sqrt{3}\times\sqrt{3})R30^{\circ}$} and p(2$\times$2) structures of fluorinated
alkanethiolate SAMs on Au(111) we have found a preference for the molecules to adsorb on bridge
sites \cite{Rusu:jpcb06}. The calculated nearest neighbor distance between the metal atoms on
Au(111) and Ag(111) is very similar, i.e., 2.94 and 2.93 \AA\ respectively. Moreover, since even in
the \mbox{$(\sqrt{7}\times\sqrt{7})$} structure the molecules show a tendency to adsorb on bridge
sites, we only consider bridge sites for the \mbox{$(\sqrt{3}\times\sqrt{3})$} and p(2$\times$2)
structures of SAMs on Ag, see Fig.~\ref{SCF3}.

Table~\ref{SCF3 and SCH2CF3} lists the molecular geometries of CF$_{3}$S and CF$_3$CH$_2$S SAMs on
Ag(111) adsorbed in bridge structures for the different packings. The geometries are in fact very
similar, with S--Au bond lengths in the range 2.5-2.6 \AA, and angles of the C--S bond with the
surface normal around 45$^\circ$. In the case of CF$_3$CH$_2$S the chain angle is $13-15^{\circ}$,
which is similar to that found in CH$_3$CH$_2$S, see Table~\ref{7times7data}.

The calculated adsorption energies indicate that a less dense packing of the fluorinated
alkanethiolate SAMs is favorable, see Table~\ref{SCF3 and SCH2CF3}. The p(2$\times$2) structure is
most stable for CF$_{3}$S, whereas for CF$_3$CH$_2$S the
\mbox{$(\sqrt{3}\times\sqrt{3})R30^{\circ}$} is slightly more stable. Some of the energy
differences are quite small, but the calculations correspond to the experimental trend observed in
fluorinated alkanethiolates on Au(111) \cite{Carla:lang93,Pflaum:surf02}.

\begin{table}[btp]
\begin{tabular}{ l  c  c  c   }
\\
\hline
 & & CF$_{3}$S &\\
\cline{2-4}
 &  $\sqrt{7}\times\sqrt{7}$   & $\sqrt{3}\times\sqrt{3}$ & $p(2\times2)$ \\
\hline
C--F(\AA) & 1.35/1.36/1.35 & 1.36 & 1.36\\
S--C(\AA) & 1.85/1.83/1.87 & 1.84 & 1.84\\
Ag--S(\AA)  & 2.51/2.41/2.57 & 2.56 & 2.54\\
C--S--normal$(^\circ)$ & 6.5/35.5/5.0 & 43.2 & 47.7\\
Ag--S--Ag$(^\circ)$ & 79/-/72 & 73.4 & 71.5 \\
$E_\mathrm{ads}$ (eV) & $-1.98$ & $-2.32$ & $-2.39$ \\
\hline
 & & CF$_3$CH$_2$S &\\
 \cline{2-4}
C--F(\AA) & 1.35 & 1.36 & 1.36\\
C--C(\AA) & 1.52 & 1.52 & 1.52\\
C--H(\AA) & 1.10 & 1.09 & 1.09\\
S--C(\AA) & 1.84 & 1.85 & 1.85\\
Ag--S(\AA)  & 2.53/2.45/2.52 & 2.56 & 2.56\\
C--C--normal$(^\circ)$ & 25.9/25.7/29.0 & 25.8 & 21.0\\
C--S--normal$(^\circ)$ & 44.0/42.4/42.7 & 42.8 & 46.7\\
chain$(^\circ)$ & 13.9/12.9/15.1 & 12.3 & 16.7\\
Ag--S--Ag$(^\circ)$ & 72.5/-/73.5 & 72.1 & 72.0 \\
$E_\mathrm{ads}$ (eV) & $-2.01$ & $-2.27$ & $-2.24$ \\

\end{tabular}
\caption{Bond lengths, bond angles and adsorption energies of CF$_{3}$S and CF$_3$CH$_2$S SAMs
adsorbed on Ag(111) surface in \mbox{$(\sqrt{7}\times\sqrt{7}) R19.1^{\circ}$},
\mbox{$(\sqrt{3}\times\sqrt{3})R30^{\circ}$} and p(2$\times$2) structures. The chain angle
represents the angle made in the CF$_3$CH$_2$S SAM by the top C atom and the sulfur atom with the
surface normal.} \label{SCF3 and SCH2CF3}
\end{table}

\subsection{Work functions and interface dipoles}

The change in work function upon adsorption of a SAM is defined by Eq.~(\ref{eq:3}). For the clean
Ag(111) surface we calculate a work function of 4.50 eV, using a $25\times 25$ \textbf{k}-point
Brillouin zone sampling grid. This is in good agreement with the experimental results of 4.5 eV
\cite{Monreal:cm03} and 4.4 eV \cite{de_Boer:adv_mat}, and with a previously reported theoretical
value of 4.42 eV, which was extracted from DFT-GGA calculations and a $15\times 15$
\textbf{k}-point sampling \cite{Bocquet:mol_phys}. The calculated work function changes upon
adsorption of the (fluorinated) alkanethiolate SAMs are given in Table~\ref{tab:Ag work_functions}.

The results clearly show that alkanethiolates decrease the work function, whereas fluorinated
alkanethiolates increase it significantly. Depending on the structure, a difference of 2-2.5 eV is
found between the work functions of alkanethiolates and partially fluorinated alkanethiolates
adsorbed on the silver surface. The work function is thus tunable over a large range by adsorption
of a suitable SAM, as is observed experimentally \cite{Campbell:prb96,de_Boer:adv_mat}. The
absolute change in the work function upon adsorption of fluorinated alkanethiolates is 3-4 times
larger than the change upon adsorption of nonfluorinated molecules. This result is quite different
from our previous findings for SAMs on Au(111), where fluorinated and nonfluorinated molecules give
a change in the work function that is similar in size (but of opposite sign, of course)
\cite{Rusu:jpcb06}. We conclude that the interaction between the molecules and the silver surface
differs from that between the molecules and the gold surface.

Kelvin probe measurements of the work function changes induced by adsorption of long-chain thiolate
SAMs on Ag(111) have been reported by Campbell \textit{et al.} \cite{Campbell:prb96}, and by de
Boer \textit{et al.} \cite{de_Boer:adv_mat}, The values reported for CH$_3$(CH$_2$)$_9$S and
CH$_3$(CH$_2$)$_{15}$S are $-0.7$ eV \cite{Campbell:prb96}, and $-0.6$ eV \cite{de_Boer:adv_mat},
respectively, and $0.9$ eV \cite{Campbell:prb96} and $1.1$ eV \cite{de_Boer:adv_mat} for
CF$_3$(CF$_2$)$_7$(CH$_2$)$_2$S. Our calculated value of $-0.59$ eV for CH$_3$CH$_2$S in the
\mbox{$(\sqrt{7}\times\sqrt{7}) R19.1^{\circ}$} bridge structure is close to the experimental
values for alkanethiolates. The calculated values for the fluorinated molecules are higher than the
experimental values even for the less densily packed p(2$\times$2) structure. One explanantion
might be that the effective dipole moment of a long-chain fluorinated molecule embedded in a SAM is
smaller than that of a short-chain molecule; in other words, the effective dielectric constant in a
long-chain fluorinated thiolate SAM is larger, see Eq.~(\ref{eq:5}). In addition, SAMs of
fluorinated alkanethiolates may show more intrinsic disorder than their nonfluorinated counterparts
\cite{Pflaum:surf02}, which also reduces the average dipole moment perpendicular to the surface.

\begin{table}[btp]
\begin{tabular}{ l  c  c  c   }
\\
\hline
           &  $ \sqrt{7}\times\sqrt{7} $  & $ \sqrt{3}\times\sqrt{3} $ &  p(2$\times$2) \\
\hline
CH$_{3}$S  & $-0.52$ ($-0.84^a$, $-0.99^b$) & $-0.61$ & $-0.34$\\
CH$_{3}$CH$_{2}$S & $-0.59$ & $-0.52$ & $-0.46$\\
CF$_{3}$S & $1.80$ ($1.79^a$) & $1.61$ & $1.48$ \\
CF$_{3}$CH$_{2}$S & $2.09$ & $1.75$ & $1.75$\\

\end{tabular}
\caption{Work functions shifts in eV with respect to the clean surface of SAMs on Ag(111) in the
bridge structure; $^{a,b}$ in the 1,3 hollow and the reconstructed structure, respectively.}
\label{tab:Ag work_functions}
\end{table}

The three $\sqrt{7}\times\sqrt{7}$ structures for CH$_3$S SAMs discussed in the previous section
give rise to different work functions. The adsorption of molecules in the bridge structure gives a
substantial smaller shift of the work function than adsorption in the 1,3 hollow structure or the
reconstructed structure. Using Eq.~(\ref{eq:3}) we can interprete the changes in the work function
in terms of molecular dipoles. This difference in work function shift between the structures can be
related to the orientation of the molecular dipoles. Whereas in the bridge structure the molecular
tails are tilted with respect to the surface normal, the tails of the two molecules that are
adsorbed at hollow sites in the 1,3 hollow structure are almost perpendicular to the surface, see
Table~\ref{7times7data}. The latter leads to larger dipole moments along the surface normal. All
the molecular tails in the reconstructed structure are perpendicular to the surface, which leads to
large dipole moments and a large work function shift. These results suggest that work function
measurements might be a simple experimental way of distinguishing between the different structures.

As can be observed in Table~\ref{tab:Ag work_functions}, the work functions have a relatively weak
dependence on the packing density. The local geometries of the thiolate molecules in the bridge
structure are similar for the different packings, see e.g. Table~\ref{SCF3 and SCH2CF3}. Hence one
would expect the individual molecular dipoles to be similar. The weak dependence of the work
functions on the packing density might seem somewhat surprising. Assuming fixed interface dipoles
$\Delta \mu$ per molecule in Eq.~(\ref{eq:3}), the work functions should scale as $1/A$, where $A$
is the surface area per molecule. Since this is clearly not the case, it means that $\Delta \mu$
depends on the packing density. In particular, the individual molecular dipoles increase with
decreasing packing density.

Decreasing the packing density increases the distances between the molecules in the SAM. Hence it
decreases the depolarizing field in the SAMs, or, in other words, the effective dielectric constant
$\varepsilon$ introduced in Eq.~(\ref{eq:5}) decreases with decreasing packing density. This
effectively increases the molecular dipoles, which opposes the effect of a decreasing density of
the molecules on the interface dipole. The net result is a weak dependence of the work function on
the packing density in the range considered and in some cases even a nonmonotonic behavior, see
Table~\ref{tab:Ag work_functions}.

In order to quantify this analysis we make use of the relations given by
Eqs.~(\ref{eq:3})-(\ref{eq:5}). We extract from the work function change an interface dipole per
molecule $\Delta \mu$ and split $\Delta \mu$ into a contribution $\mu_\mathrm{SAM}$ from the
molecular dipole and a contribution $\mu_\mathrm{chem}$ from the charge transfer between the
molecule and the surface upon chemisorption. The results for each of the SAMs in the
$\sqrt{7}\times\sqrt{7}$, $\sqrt{3}\times\sqrt{3}$ and p(2$\times$2) structures are reported in
Table~\ref{table:dip}. As we are explicitly interested in the influence of the packing density we
continue to compare similar, i.e. bridge-like, structures.

$\mu_\mathrm{SAM}$ is positive for fluorinated alkanethiolates, which means that the molecular
dipoles point from the S atom to the CF$_3$ group. For the nonfluorinated alkanethiolates the
absolute values of $\mu_\mathrm{SAM}$ are larger, but the sign is negative, meaning that the
dipoles point from the alkyl tails to the S atom. $\mu_\mathrm{chem}$ is positive for all
molecules. The latter contribution is associated with a dipole that points from the surface to the
molecule. It is associated with a (partial) electron transfer from the surface to the molecule.
Both contributions, $\mu_\mathrm{SAM}$ and $\mu_\mathrm{chem}$, to the interface dipole $\Delta
\mu$ are of comparable size. For the nonfluorinated molecules they are of opposite sign, which
leads to moderate interface dipoles $\Delta \mu = -0.2$ to $-0.3$ D and work function changes
$\lesssim 0.5$ eV. The contributions $\mu_\mathrm{SAM}$ and $\mu_\mathrm{chem}$ have the same sign
for the fluorinated molecules, which gives large interface dipoles $\Delta \mu = 0.9$-$1.3$ D and
large work function changes of up to 2-2.5 eV.

\begin{table*}[!tbp]

\begin{tabular}{lcccc|cccc|cccc}
\\
\hline
&\multicolumn{4}{c}{$\sqrt{7}\times\sqrt{7}$}  &\multicolumn{4}{c}{$\sqrt{3}\times\sqrt{3}$}  &\multicolumn{4}{c}{p(2$\times$2)} \\
\cline{2-5}\cline{6-9}\cline{10-13} & CH$_3$S & C$_2$H$_5$S & CF$_3$S & CF$_3$CH$_2$S & CH$_3$S &
C$_2$H$_5$S & CF$_3$S &
CF$_3$CH$_2$S & CH$_3$S & C$_2$H$_5$S & CF$_3$S & CF$_3$CH$_2$S\\
\hline
$\Delta\mu        $   &$-0.24$  & $-0.27$ & 0.83   & $0.96$     & $-0.36$ & $-0.31$ & $0.95$  & $1.04$ &   $-0.27$ & $-0.36$ &$1.17$  & $1.38$\\
$\mu_\mathrm{SAM} $   &$-0.78$  & $-0.79$ & 0.39   & $0.43$     & $-0.88$ & $-0.79$ & $0.44$  & $0.50$ &   $-0.92$ & $-1.01$ &$0.50$  & $0.71$\\
$\mu_\mathrm{chem}$   &$0.54$   &  $0.52$ & 0.44   & $0.53$     &  $0.52$ &  $0.46$ & $0.51$  & $0.54$ &   $0.65$  & $ 0.65$ &$0.67$  & $0.67$\\
$\varepsilon$         & 1.6    &    1.4   & 2.0    &  2.7       &   1.4   &   1.4   &  1.8    &  2.3   &    1.3    & 1.1     & 1.6    &  1.7  \\
\end{tabular}
\caption{Dipole per molecule $\Delta\mu$ from work function shift upon adsorption, the
(perpendicular) molecular dipole moment $\mu_\mathrm{SAM}$ in a free standing SAM and the
chemisorption dipole moment $\mu_\mathrm{chem}$ of the SAMs on Ag(111) surface. The values are in
D. $\varepsilon$ is the effective dielectric constant of the free standing SAM.\label{table:dip}}
\end{table*}

\begin{figure*}[!tbp]

\includegraphics[scale=0.68,clip=true]{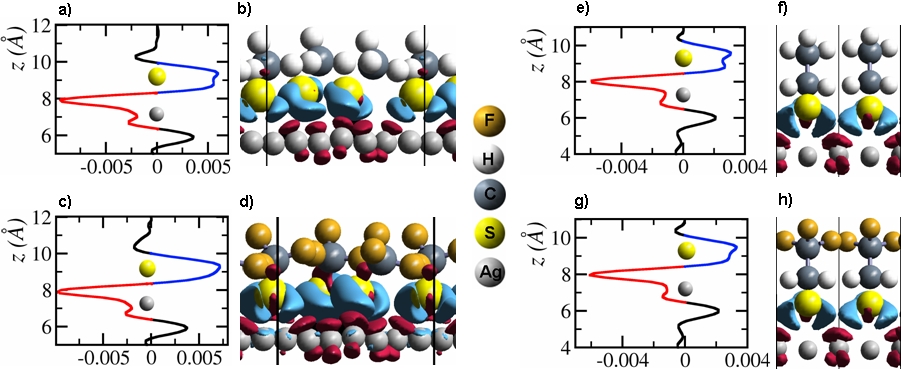}
\caption{Difference electron density along the surface normal ($z$) averaged over the $xy$ plane in
unites of \AA$^{-3}$ and as isodensity surface. (a), (b) CH$_{3}$S and (c), (d) and CF$_{3}$S in
the $\sqrt{7}\times\sqrt{7}$ structure; (e), (f) CH$_{3}$CH$_{2}$S and (g), (h) CF$_{3}$CH$_{2}$S
in the p(2$\times$2) structure.}\label{charge}

\end{figure*}

Comparing the results for the different packing densities in Table~\ref{table:dip}, we observe that
the absolute value of $\mu_{SAM}$ increases if the packing density decreases. This can be
understood by noting that the effective dielectric constant $\varepsilon$ of the SAM decreases if
the packing density decreases, as discussed above. We have calculated the dipole moment of the
isolated alkanethiolate radical molecules, fixing the molecules in the geometries they attain in
the SAM. The component $\mu_z$ along the surface normal in the adsorbed geometry is given in
Table~\ref{table:electronegativities}. From Eq.~(\ref{eq:5}) we then calculate the effective
dielectric constant of the free standing SAM. The dielectric constants for fluorinated
alkanethiolate SAMs are somewhat larger than those of their nonfluorinated counterparts. This might
be expected since the polarizability of fluorinated molecules is larger and therefore screening in
the SAM is larger. The results in Table~\ref{table:dip} clearly show that the dielectric constants
decrease with decreasing packing density.

$\mu_\mathrm{chem}$ also increases with decreasing packing density. The origin of this effect is
similar to that discussed in the previous paragraph; the screening of the dipoles in the layer
decreases if the packing density decreases. An interesting observation is that at fixed packing
density $\mu_\mathrm{chem}$ shows little variation within the range of molecules. Apparently it is
mainly determined by the S--Ag interaction and not so much by the molecular tails.

Upon adsorption electronic charge is transferred from the surface to the molecule. In order to
visualize the charge transfer at the interface upon adsorption of the SAMs, we calculate the
change in electron density $\Delta n$:
\begin{equation}
\Delta n = n_\mathrm{SAM-Ag}-n_\mathrm{Ag}-n_\mathrm{SAM}, \label{eq:7}
\end{equation}
where $n_\mathrm{SAM-Ag}$, $n_\mathrm{Ag}$ and $n_\mathrm{SAM}$ are the electron densities of the
SAM adsorbed on Ag(111), of the Ag(111) surface and of the free-standing SAM, respectively. The
electron distributions are obtained on a real space grid from separate calculations on the adsorbed
SAM, on the substrate and on the free-standing SAM, respectively. In the latter two calculations
the substrate and the molecules are fixed in the adsorption geometries.

As examples, Fig.~\ref{charge} shows the difference electron density $\Delta n$ averaged in the
$xy$ plane along the surface normal of CH$_{3}$S and CF$_{3}$S SAMs adsorbed in the
$\sqrt{7}\times\sqrt{7}$ structure and of CH$_{3}$CH$_{2}$S and CF$_{3}$CH$_{2}$S SAMs adsorbed in
the p(2$\times$2) structure. In addition three dimensional visualizations of $\Delta n$ at the
interface are presented. Only the region around the SAM/Ag(111) interface is shown, since in the
substrate and the vacuum region $\Delta n$ $\simeq$ 0. The figures clearly demonstrate that $\Delta
n$ is localized at the interface, i.e., near the sulfur atoms and the top metal layer. Electronic
density is transferred mainly from the top layer of silver atoms to the sulfur atoms, which results
in a dipole moment $\mu_\mathrm{chem}$.

To check the consistency of this analysis, we can calculate the metal-sulfur dipoles from the
difference electron density:
\begin{equation}
\mu'_\mathrm{chem}=\int\int\int_{z_{o}}^{z_{v}} z\Delta n(x,y,z)dxdydz, \label{eq:8}
\end{equation}
where we choose $z_{o}$ in the center between the second and the third metal layer of the substrate
and $z_{v}$ in the center of the vacuum. The metal-sulfur dipoles $\mu'_\mathrm{chem}$ are within
10\% of the values $\mu_\mathrm{chem}$ listed in Table~\ref{table:dip}.

By integrating the peak of $\Delta n$ on the sulfur atom, see Fig.~\ref{charge}, one can calculate
the charge transfer from the substrate to the sulfur atom. A typical value over a range of
structures is $q=(0.11\pm0.01)e$. Modeling the charge transfer dipole as $\mu_{chem}=qd$ gives
$d=1.1$\AA, assuming $\mu_\mathrm{chem}=0.6$ D. The distance between the sulfur atoms and the top
layer of silver atoms is 2.0-2.2 \AA , so this analysis is consistent with the interpretation of
$\mu_\mathrm{chem}$ as a metal-sulfur bond dipole.

The fact that this dipole moment hardly depends on the different molecular tails is slightly
surprising, since the electronegativity of fluorinated tails is much higher than that of
unfluorinated ones. The Mulliken electronegativity of a molecule is defined as:
\begin{equation}
\chi_{M}= \frac{EA+IP}{2}, \label{eq:9}
\end{equation}
where $EA$ and $IP$ are the electron affinity and ionization potential of the molecule. The $EA$s,
$IP$s and Mulliken electronegativities of the molecules considered in this paper (fixed in their
adsorbed geometries) are given in Table~\ref{table:electronegativities}.

One observes a considerable difference in the electronegativities $\chi_M$ of the molecules. The
$\chi_M$ of both alkanethiolates is similar, but the $\chi_M$ of fluorinated alkanethiolates is
much larger. The HOMO of the radical neutral molecules, which plays a role in determining the $EA$
and $IP$, is stabilized by the electron withdrawing CF$_{3}$ group. This property is commonly
associated with the attractive Coulomb field of the CF$_{3}$ group. The HOMO of the neutral
molecules is localized mainly on the sulfur atom and one expects that the effect of the CF$_{3}$
group decreases if the distance between this group and the sulfur atom increases. Indeed one finds
that the $EA$, $IP$ and $\chi_M$ of CF$_3$S are significantly higher that those of CF$_3$CH$_2$S.

The electronegativity of a metal surface is given by its work function. From simple chemical
reasoning one would assume that the charge transfer between a molecule and a surface would depend
on the difference of their electronegativities. This is clearly not the case; the electronegativity
of the molecule does not seem to influence the charge transfer. This suggests that the effects of
the Coulomb field of the CF$_{3}$ group and the alkyl tails on the charge distribution at the
sulfur-metal interface are screened by the metal.

\begin{table}[!]
\begin{tabular}{lcccc}
\\
\hline
& CH$_3$S & C$_2$H$_5$S & CF$_3$S & CF$_3$CH$_2$S \\
\hline
$EA$                 &  1.73  &  1.87  &  3.02   &  2.43  \\
$IP$                 &  9.20  &  8.95  & 10.79   &  9.82  \\
$\chi_M$             &  5.47  &  5.41  &  6.91   &  6.13  \\
$|\mu_\mathrm{tot}|$ &  1.70  & 1.82   &  1.05   &  2.06  \\
$\mu_z$              &$-1.23$ &$-1.14$ &  0.79   &  1.17
\end{tabular}
\caption{Electron affinity (EA), ionization potential (IP), Mulliken electronegativity ($\chi_M$)
in eV, total dipole moment $|\mu_\mathrm{tot}|$ and dipole moment along the surface normal $\mu_z$
in D of isolated molecules in their adsorbed geometries. \label{table:electronegativities}}
\end{table}

\section{Summary and conclusions}

We have studied the interface dipole formation and work function changes produced by adsorption of
CH$_{3}$S, CH$_{3}$CH$_{2}$S, CF$_{3}$S and CF$_{3}$CH$_{2}$S SAMs on the Ag(111) surface by means
of DFT calculations. Adsorption of the alkanethiolates CH$_{3}$S and CH$_{3}$CH$_{2}$S decreases
the work function as compared to the clean metal surface, whereas adsorption of the fluorinated
alkanethiolates CF$_{3}$S and CF$_{3}$CH$_{2}$S increases the work function.

In particular we have examined the influence of the structure and the packing density of the
molecules in the SAM. CH$_{3}$S on Ag(111) in the unreconstructed \mbox{$(\sqrt{7}\times\sqrt{7})
R19.1^{\circ}$} structure with two of the three molecules in the unit cell adsorbed at a hollow
site, leads to a work function shift of $-0.8$ eV. Adsorbing the CH$_{3}$S molecules on bridge
sites stabilizes the structure by 0.10 eV/molecule and gives a work function shift of $-0.5$ eV.
The recently proposed surface reconstruction induced by CH$_{3}$S adsorption yields an almost
identical adsorption energy, and a work function shift of $-1.0$ eV. The difference between the
work functions of these structures can be interpreted in terms of the difference in the orientation
of the molecular dipoles.

These results are compared to the less densily packed \mbox{$(\sqrt{3}\times\sqrt{3}) R30^{\circ}$}
and \mbox{p(2$\times$2)} structures, which are more likely to occur for fluorinated alkanethiolate
SAMs. Although the work function shift generally decreases for decreasing packing density, it is
not simply proportional to the density of molecular dipoles. A partial compensating effect is
caused by a decrease of the dielectric screening in the molecular layer.

Comparing the different molecules adsorbed in similar geometries shows that fluorinated
alkanethiolates can increase the work function by up to 2 eV, which is much larger that the
decrease in work function caused by (nonfluorinated) alkanethiolate adsorption. We explain this by
separating the interface dipole into a contribution from the molecular dipoles and from the charge
reordering at the metal$-$SAM interface. Electron transfer occurs from the Ag surface to the sulfur
atoms of the thiolate molecules. The resulting dipole points in the same direction as the molecular
dipole for fluorinated molecules. Addition of the two dipoles leads to a large interface dipole and
a large work function shift. The direction of the molecular dipole of nonfluorinated molecules is
opposite to the metal-sulfur bond dipole, resulting in a much smaller interface dipole and work
function shift.

The electron transfer from the Ag surface to the molecules is remarkably independent of the
electronegativity of the molecules. In good approximation the charge reordering only depends upon
the metal-sulfur bond, which suggest that the influence of the molecular tails is screened by the
metal substrate. In previous calculations we arrived at the same conclusion for adsorption of
alkanethiolate SAMs on other noble metal surfaces, indicating that this result is more general
\cite{Rusu:jpcb06,Rusu:prb06}. For adsorption on Ag(111) we find effective Ag-S dipoles
$\mu_\mathrm{chem} = 0.51 \pm 0.04$ D and $0.66 \pm 0.01$ D in the
\mbox{$(\sqrt{3}\times\sqrt{3})$} and \mbox{p(2$\times$2)} structures, respectively.

For adsorption on Au(111) we have found very small Au-S dipoles $\mu_\mathrm{chem} < 0.1$ D,
indicating an apolar Au-S bond \cite{Rusu:jpcb06}, whereas for adsorption on Pt(111) the Pt-S
dipole is $\mu_\mathrm{chem} = -0.45\pm 0.03$ D \cite{Rusu:prb06}. The latter indicates an electron
transfer from the sulfur atoms to the Pt surface. The metal-sulfur bonds formed upon SAM adsorption
generate dipole moments that reduce the work function differences between the clean metal surfaces,
and can even reverse the order. The molecular dipole moments exhibit similar values for adsorption
on Ag, Au and Pt, thus giving the possibility to design interface dipoles. By adding molecular and
metal-sulfur bond dipoles the overall work function can be determined. It is possible to manipulate
metal work functions considerably using SAMs. Work function shifts that can be as large as 2 eV.

\begin{acknowledgments}
This work is supported by ``Prioriteits Programma Materialenonderzoek" (PPM), by the ``Stichting
voor Fundamenteel Onderzoek der Materie" (FOM), financially supported by the ``Nederlandse
Organisatie voor Wetenschappelijk Onderzoek" (NWO), and by ``NanoNed'', a nanotechnology program of
the Dutch Ministry of Economic Affairs. The use of supercomputer facilities was sponsored by the
``Stichting Nationale Computer Faciliteiten" (NCF), financially supported by NWO.
\end{acknowledgments}


\end{document}